\documentclass[runningheads]{llncs}
\usepackage{tikz}
\usetikzlibrary{shapes.geometric}
\usepackage{pgfbaseshapes}
\usepackage{graphicx}
\usepackage{pgfplots}
\usepackage{floatrow}
\usepackage[linesnumbered,ruled,vlined]{algorithm2e}
\SetAlFnt{\scriptsize}
\SetAlCapFnt{\scriptsize}
\SetAlCapNameFnt{\scriptsize}

\usepackage{amsmath}
\usepackage{color}
\usepackage{xspace}
\usepackage{hyperref}
\usepackage{url}
\usepackage{subcaption}
\usepackage{listings}
\usepackage{lmodern}
\usepackage{calc}
\usepackage{ifthen}
\usepackage[iso,english]{isodate}
\usepackage{marginnote}
\usepackage[scaled=0.88]{helvet}
\usepackage{pgf,tikz}
\usepackage{amssymb}
\usepackage{lipsum}
\usepackage{cite}
\usepackage[smaller,nolist,nohyperlinks]{acronym}
\newcommand{\acposs}[1]{%
	\acs{#1}'s%
}
\usepackage{sidecap}
\usepackage{wrapfig}
\usepackage{rotating}
\usepackage[T1]{fontenc}
\usepackage{romannum}
\usepackage{pdfpages}
\usepackage{csquotes}

\DeclareGraphicsExtensions{.pdf,.jpeg,.png}
\graphicspath{{images/}}
\usetikzlibrary{circuits}

\newboolean{anonymous} % anonymous version
\setboolean{anonymous}{false}
\newcommand{\onlyan}[1]{\ifthenelse{\boolean{anonymous}}{#1}{}}
\newcommand{\onlynonan}[1]{\ifthenelse{\boolean{anonymous}}{}{#1}}

\newlength{\onecollinewidth}
\newcounter{hours}
\newcounter{minutes}
\newcommand{\printtime}{%
  \setcounter{hours}{\time/60}%
  \setcounter{minutes}{\time-\value{hours}*60}%
  \thehours :\ifthenelse{\theminutes < 10}{0}{}\theminutes}

\lstdefinelanguage{SCL}
{language=C,
morekeywords={bool, end, false, fork, fork1, input, join, join1, module,
  output, pause, par, then, tickstart, tickreturn, true}}

\lstdefinelanguage{XTEND}
{language=C,
  morekeywords={def, if, null, void, for, val, var, nullOrEmpty, immutableCopy, immutableCopy,     setTypeConnector, id, setImmediate, addEffect, setTypeNormal},
  deletekeywords={const},
  moredelim=**[is][\color{red}]{@}{@},
  breaklines=true,  
  tabsize=2,
  breakindent=2mm,
  breakatwhitespace=true,
  basicstyle=\sffamily,
  columns=fullflexible,
  backgroundcolor=\color{gray!10}
}

\lstdefinelanguage{KICO}
{language=C,
  morekeywords={def, if, null, void, for, val, var, nullOrEmpty, immutableCopy, immutableCopy,     setTypeConnector, id, setImmediate, addEffect, setTypeNormal, system, pre, post, set, process, label, public},
  deletekeywords={const},
  moredelim=**[is][\color{red}]{@}{@},
  breaklines=true,  
  tabsize=2,
  breakindent=2mm,
  breakatwhitespace=true,
  basicstyle=\sffamily,
  columns=fullflexible,
  backgroundcolor=\color{gray!10}
}

\newcommand{\textsff}[1]{\textsf{#1}}

\newcommand{\Eg}{E.\,g.\@\xspace}
\newcommand{\eg}{e.\,g.\@\xspace}

\newcommand{\ie}{i.\,e.\@\xspace}
\newcommand{\etal}{et al.\@\xspace}

\newcommand{\node}{\mathit{node}}

\graphicspath{{images/}}
\setboolean{anonymous}{false}

\begin{acronym}[KIELER]	
	\acro{api}[API]{Application Programming Interface}
	\acro{css}[CSS]{Cascading Style Sheets}
	\acro{dom}[DOM]{Document Object Model}
	\acro{dsl}[DSL]{Domain Specific Language}
	\acro{elk}[ELK]{Eclipse Layout Kernel}
	\acro{elklayered}[ELK Layered]{\acs{elk} Layered}
	\acro{emf}[EMF]{Eclipse Modeling Framework}
	\acro{gwt}[GWT]{Google Web Toolkit}
	\acro{hci}[HCI]{Human Computer Interaction}
	\acro{html}[HTML]{Hypertext Markup Language}
	\acro{ide}[IDE]{Integrated Development Environment}
	\acro{iur}[iur]{initialize-update-read}
	\acro{j2ee}[J2EE]{Java 2 Platform, Enterprise Edition}
	\acro{jdk}[JDK]{Java Development Kit}
	\acro{jdt}[JDT]{Eclipse Java development tools}
	\acro{json}[JSON]{JavaScript Object Notation}
	\acro{json-rpc}[JSON-RPC]{\acs{json} Remote Procedure Call}
	\acro{jvm}[JVM]{Java Virtual Machine}
	\acro{keith}[KEITH]{\acs{kiel} Environment Integrated in Theia}
	\acro{kico}[KiCo]{\acs{kieler} Compiler}
	\acro{kiel}[KIEL]{Kiel Integrated Environment for Layout}
	\acro{kieler}[KIELER]{Kiel Integrated Environment for Layout Eclipse Rich Client}
	\acro{kiml}[KIML]{Kieler Infrastructure for Meta Layout}
	\acro{klighd}[KLighD]{\acs{kieler} Lightweight Diagrams}
	\acro{klodd}[KLoDD]{\acs{kieler} Layout of Dataflow Diagrams}
	\acro{klay}[KLay]{\acs{kieler} Layout}
	\acro{lsp}[LSP]{Language Server Protocol}
	\acro{mde}[MDE]{Model-Driven Engineering}
	\acro{os}[OS]{Operating System}
	\acro{osgi}[OSGi]{Open Service Gateway Initiative}
	\acro{pde}[PDE]{Plug-in Development Environment}
	\acro{rca}[RCA]{Rich Client Application}
	\acro{sccharts}[SCCharts]{Sequential Constructive Statecharts}
	\acro{scl}[SCL]{Sequentially Constructive Language}
	\acro{svg}[SVG]{Scalable Vector Graphic}
	\acro{swt}[SWT]{Standard Widget Toolkit}
	\acro{tcp}[TCP]{Transmission Control Protocol}
	\acro{ui}[UI]{user interface}
	\acro{uri}[URI]{Uniform Resource Identifier}
	\acro{url}[URL]{Uniform Resource Locator}
	\acro{ux}[UX]{user experience}
	\acro{vscode}[VSCode]{Visual Studio Code}
	\acro{xml}[XML]{Extensible Markup Language}
	\acro{yang}[YANG]{Yet Another Next Generation}
	\acrodef{RiCharts}{for Rich Charts}
	\end{acronym}
	
\begin{document}
\title{The Eclipse Layout Kernel}
\onlynonan{
	\author{S\"oren Domr\"os\inst{1}\orcidID{0000-0002-8011-8484} \and
		Reinhard~von~Hanxleden\inst{1}\orcidID{0000-0001-5691-1215} \and
		Miro Sp\"onemann\inst{2} \and
		Ulf R\"uegg\inst{1} \and
		Christoph Daniel Schulze\inst{1}
	}
	\authorrunning{S. Domr\"os et al.}
	\institute{
		Kiel University  \email{\{sdo, rvh\}@informatik.uni-kiel.de, uruurumail@gmail.com, cd.schulze@gmx.net}	\and
		Typefox GmbH \email{miro.spoenemann@typefox.io}
	}
}
\maketitle
\begin{abstract}
The \ac{elk} is a collection of graph drawing algorithms that supports compound graph layout and ports as explicit anchor points of edges.
It is available as open-source library under an EPL license.
Since its beginning, \ac{elk} has served both as a research vehicle for graph drawing algorithms, and as a practical tool for solving real-world problems.
\ac{elk} and its transpiled JavaScript cousin elkjs are now included in numerous academic and commercial projects.

Most of the algorithms realized in \ac{elk} are described in a series of publications.
In this paper, the technical description concentrates on the key features of the flag-ship algorithm \acs{elk} Layered, the algorithm architecture, and usage.
However, the main purpose of this paper is to give the broader view that is typically left unpublished.
Specifically, we review its history, give a brief overview of technical papers, discuss lessons learned over the past fifteen years, and present example usages.
Finally, we reflect on potential threats to open-source graph drawing libraries.

\end{abstract}
\acresetall
\acused{ide}
\acused{dsl}
\acused{ui}
\acused{rca}
\section{Introduction}
\label{sec:introduction}
\pagenumbering{arabic}
\setcounter{footnote}{0}

Freely available, high-quality graph drawing libraries are useful both for stress-testing their underlying algorithms, and for harnessing the power of automatic graph drawing in practical applications.
In 2004, Jünger and Mutzel published a volume of papers on various graph drawing libraries~\cite{JuengerM03}.
This paper presents the \ac{elk}, which has been developed since 2008 and is still actively maintained, with a growing user base.

The original driver of the \ac{elk} is modeling pragmatics \cite{vonHanxledenLF+22}, which, in a nutshell, aims to increase the productivity of designers working with graphical modeling languages.
This is still the main focus of \ac{elk}, and explains why modeling languages such as SCCharts \cite{vonHanxledenDM+14}, Ptolemy \cite{Ptolemaeus14}, or, more recently, Lingua Franca \cite{LohstrohMBL21}, feature prominently in publications on \ac{elk}.
In fact, when getting started on modeling pragmatics with the \ac{kiel} tool \cite{ProchnowT05} for model-driven engineering of statecharts \cite{Harel87}, the developers started with the off-the-shelf layout library Graphviz \cite{GansnerN00}.
It was only when they noticed that existing libraries did not fully address their needs that they started developing and publishing their own algorithms.
We believe this practical motivation, and continuously tight interaction between researchers and practitioners of model-driven engineering, has contributed to the quality and usability of \ac{elk}.

\autoref{fig:timeline} displays \acposs{elk} development timeline with selected student theses\footnote{\url{https://www.rtsys.informatik.uni-kiel.de/en/teaching/theses/completed-theses}}.
The diploma thesis of Spönemann led to the first publication of the \ac{klodd} algorithm to visualize data-flow \cite{SpoenemannFvH+09}.
That thesis subsequently led to the development of the \ac{elk} predecessor \ac{klay}, which already included a predecessor of \acp{elk} flagship algorithm, which was at that time called \ac{klay} Layered \cite{SchulzeSvH14}.
Each transformation step from \ac{klodd} to \ac{elk} as well as each \ac{elk} release marks not only new algorithms and features but also big refactoring efforts, which enabled the longevity of \ac{elk}.

\begin{figure}
	\centering
	\includegraphics[scale=0.1491]{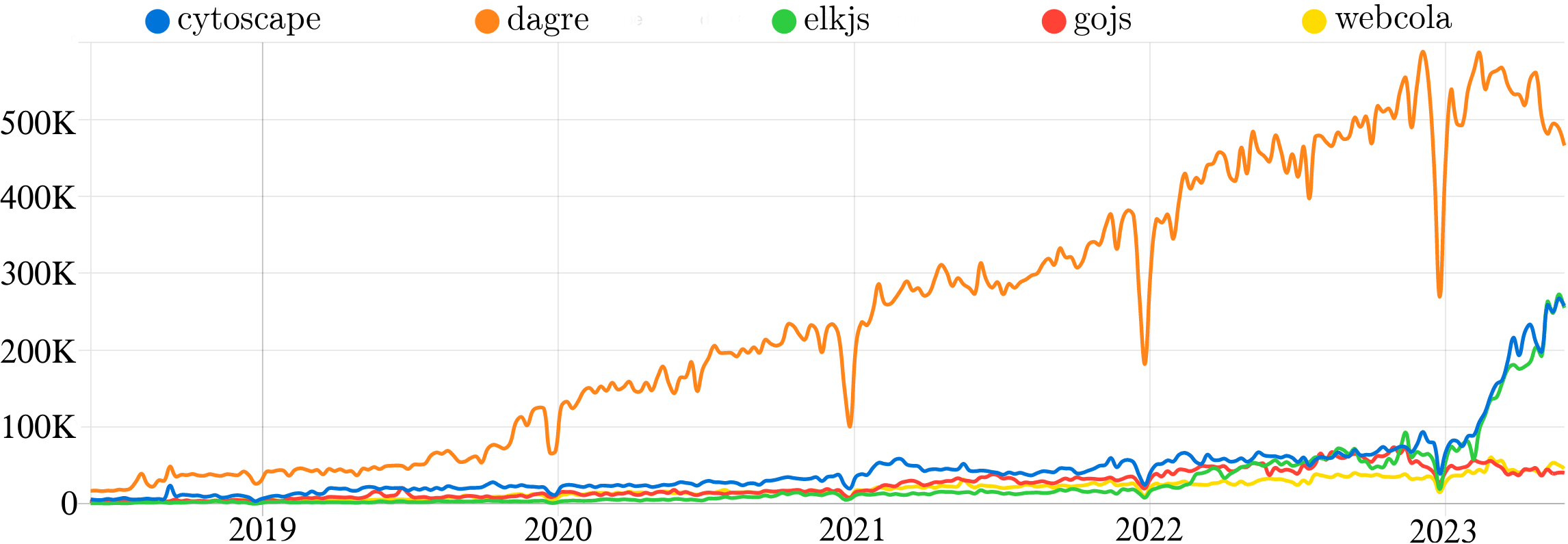}
	\caption{Weekly downloads of elkjs (green) compared to other popular layout libraries tagged with layout, graph, or dataflow on npm (\url{https://npmtrends.com/cytoscape-vs-dagre-vs-elkjs-vs-gojs-vs-webcola}, retrieved: 05/06/2023).
	% The most downloaded software dagre\protect\footnotemark (blue) is no longer maintained.
%  \rsdo{I think we don't need this image}
	}
	\label{fig:npm-trends}
\end{figure}

\definecolor{gold}{RGB}{0, 100, 0}%
\definecolor{goldold}{RGB}{255, 191, 0}%
\definecolor{gray}{RGB}{76, 76, 76}%
\definecolor{timeLineBG}{RGB}{200, 200, 200}%+
\def\timelineheight{1.5}%
\def\tlsmallheight{0.75}%
\def\startYOne{1}%
\def\startYTwo{4}%
\def\startYThree{7}%
\def\startYFour{10}%
\def\bottomMargin{-0.1}%
\def\leftMargin{0.2}%
\def\padding{0.12}%
\def\fontheight{0.21}%
\def\elkheight{0.1}%
\def\etheight{0.1}%
\def\etbmargin{0.05}%
\def\etwidth{0.1}%
\def\etlmargin{0.15}%
\def\elheight{1.2}%
\def\monthWidth{0.25}%
\def\thHeighth{0.08}%
\def\thlineheight{0.2}%
\def\thlinethick{1}%
\def\thLabelTopMargin{0.25}%
\def\thLabelSecondHeight{0.5}%
\def\thLabelHeight{0.25}%
\noindent %
\begin{figure}%
\begin{tikzpicture}[scale=0.99]%
{%
	\hypersetup{
		colorlinks=false,
		pdfborder={0 0 0},
	}%
	% Timeline 2021 - 2024
	\fill[fill=timeLineBG, fill opacity=0.1] (0, \startYOne) rectangle ++ (12,\timelineheight);
	\draw[fill=timeLineBG, opacity=0.2] (0, \startYOne) -- (0, \startYOne + \timelineheight);
	\draw[fill=timeLineBG, opacity=0.2] (0.5, \startYOne) -- (0.5, \startYOne + \tlsmallheight);
	\draw[fill=timeLineBG, opacity=0.2] (1, \startYOne) -- (1,  \startYOne + \tlsmallheight);
	\draw[fill=timeLineBG, opacity=0.2] (1.5, \startYOne) -- (1.5,  \startYOne + \tlsmallheight);
	\draw[fill=timeLineBG, opacity=0.2] (2, \startYOne) -- (2,  \startYOne + \tlsmallheight);
	\draw[fill=timeLineBG, opacity=0.2] (2.5, \startYOne) -- (2.5,  \startYOne + \tlsmallheight);
	\draw[fill=timeLineBG, opacity=0.2] (3, \startYOne) -- (3, \startYOne + \timelineheight);
	\draw[fill=timeLineBG, opacity=0.2] (3.5, \startYOne) -- (3.5, \startYOne + \tlsmallheight);
	\draw[fill=timeLineBG, opacity=0.2] (4, \startYOne) -- (4,  \startYOne + \tlsmallheight);
	\draw[fill=timeLineBG, opacity=0.2] (4.5, \startYOne) -- (4.5,  \startYOne + \tlsmallheight);
	\draw[fill=timeLineBG, opacity=0.2] (5, \startYOne) -- (5,  \startYOne + \tlsmallheight);
	\draw[fill=timeLineBG, opacity=0.2] (5.5, \startYOne) -- (5.5,  \startYOne + \tlsmallheight);
	\draw[fill=timeLineBG, opacity=0.2] (6, \startYOne) -- (6, \startYOne + \timelineheight);
	\draw[fill=timeLineBG, opacity=0.2] (6.5, \startYOne) -- (6.5, \startYOne + \tlsmallheight);
	\draw[fill=timeLineBG, opacity=0.2] (7, \startYOne) -- (7,  \startYOne + \tlsmallheight);
	\draw[fill=timeLineBG, opacity=0.2] (7.5, \startYOne) -- (7.5,  \startYOne + \tlsmallheight);
	\draw[fill=timeLineBG, opacity=0.2] (8, \startYOne) -- (8,  \startYOne + \tlsmallheight);
	\draw[fill=timeLineBG, opacity=0.2] (8.5, \startYOne) -- (8.5,  \startYOne + \tlsmallheight);
	\draw[fill=timeLineBG, opacity=0.2] (9, \startYOne) -- (9, \startYOne + \timelineheight);
	\draw[fill=timeLineBG, opacity=0.2] (9.5, \startYOne) -- (9.5, \startYOne + \tlsmallheight);
	\draw[fill=timeLineBG, opacity=0.2] (10, \startYOne) -- (10,  \startYOne + \tlsmallheight);
	\draw[fill=timeLineBG, opacity=0.2] (10.5, \startYOne) -- (10.5,  \startYOne + \tlsmallheight);
	\draw[fill=timeLineBG, opacity=0.2] (11, \startYOne) -- (11,  \startYOne + \tlsmallheight);
	\draw[fill=timeLineBG, opacity=0.2] (11.5, \startYOne) -- (11.5,  \startYOne + \tlsmallheight);
	\draw[fill=timeLineBG, opacity=0.2] (12, \startYOne) -- (12, \startYOne + \timelineheight);
	% Year labels
	\node (Year2020) at (1.5, \startYOne + \timelineheight) [above, color=gray] {2020};
	\node (Year2021) at (4.5, \startYOne + \timelineheight) [above, color=gray] {2021};
	\node (Year2022) at (7.5, \startYOne + \timelineheight) [above, color=gray] {2022};
	\node (Year2023) at (10.5, \startYOne + \timelineheight) [above, color=gray] {2023};
	% Pubs
	% 2021
	\node (DomroesLvHJ21) at (3 + \leftMargin, \startYOne + \bottomMargin) [above, color=gray] {\cite{DomroesLvHJ21}};
	% 2022
	\node (DomroesvH22) at (6 + \leftMargin, \startYOne + \bottomMargin) [above, color=gray] {\cite{DomroesvH22}};
	% 2023
	\node (DomroesLvHJ23) at (9 + \leftMargin, \startYOne + \bottomMargin) [above, color=gray] {\cite{DomroesLvHJ23}};
	\node (DomroesLvHJ23) at (9 + \leftMargin, \startYOne + \padding + \fontheight + \bottomMargin) [above, color=gray] {\cite{DomroesRvH23}};
	\node (PetzoldDSvH23) at (9 + \leftMargin, \startYOne + \padding + \padding + \fontheight + \fontheight + \bottomMargin) [above, color=gray] {\cite{PetzoldDSvH23}};
	% ELK triangles
	% 2020
	\fill[above, fill=gray] (1.5 + \etlmargin, \startYOne + \etbmargin) node[anchor=north]{}
	-- (1.5 + \etlmargin + \etwidth, \startYOne + \etheight + \etbmargin) node[anchor=north]{}
	-- (1.5 + \etlmargin + \etwidth + \etwidth, \startYOne + \etbmargin) node[anchor=south]{}
	-- cycle;
	% 2022
	\fill[above, fill=gray] (6.5 + \etlmargin, \startYOne + \etbmargin) node[anchor=north]{}
	-- (6.5 + \etlmargin + \etwidth, \startYOne + \etheight + \etbmargin) node[anchor=north]{}
	-- (6.5 + \etlmargin + \etwidth + \etwidth, \startYOne + \etbmargin) node[anchor=south]{}
	-- cycle;
	% ELK Labels
	% 2020
	\node (ELK070) at (1.75, \startYOne + \elheight) [text centered, color=gray] {0.7.0};
	% 2022
	\node (ELK080) at (6.75, \startYOne + \elheight) [text centered, color=gray] {0.8.0};
	% 2020
	% BT Carstensen
	\draw[color=gold, line width=\thlinethick] (0.75, \startYOne - \thHeighth) -- (0.75, \startYOne + \thHeighth);\thHeighth
	\draw[color=gold, line width=\thlinethick] (0.75, \startYOne) -- (0.75 + \monthWidth * 6, \startYOne);\thHeighth
	\draw[color=gold, line width=\thlinethick] (0.75 + \monthWidth * 6, \startYOne - \thHeighth) -- (0.75 + \monthWidth * 6, \startYOne + \thHeighth);
	\node (BTCarstensen) at (1.5, \startYOne - \thLabelHeight) [text centered, color=gold] {\scriptsize BT Carstensen};
	% 2021
	% MT Kasperowski
	\draw[color=gold, line width=\thlinethick] (4, \startYOne - \thHeighth) -- (4, \startYOne + \thHeighth);\thHeighth
	\draw[color=gold, line width=\thlinethick] (4, \startYOne) -- (4 + \monthWidth * 6, \startYOne);\thHeighth
	\draw[color=gold, line width=\thlinethick] (4 + \monthWidth * 6, \startYOne - \thHeighth) -- (4 + \monthWidth * 6, \startYOne + \thHeighth);
	\node (MTKasperowski) at (4.1, \startYOne - \thLabelHeight) [text centered, color=gold] {\scriptsize MT Kasperowski};
	% 2021 - 2022
	% BT Riepe
	\draw[color=gold, line width=\thlinethick] (5.25, \startYOne - \thHeighth - \thlineheight) -- (5.25, \startYOne + \thHeighth - \thlineheight);\thHeighth
	\draw[color=gold, line width=\thlinethick] (5.25, \startYOne - \thlineheight) -- (5.25 + \monthWidth * 6, \startYOne - \thlineheight);\thHeighth
	\draw[color=gold, line width=\thlinethick] (5.25 + \monthWidth * 6, \startYOne - \thHeighth - \thlineheight) -- (5.25 + \monthWidth * 6, \startYOne + \thHeighth - \thlineheight);
	\node (BTRiepe) at (6, \startYOne - \thlineheight - \thLabelHeight) [text centered, color=gold] {\scriptsize BT Riepe};
	\fill[fill=timeLineBG, fill opacity=0.1] (0, \startYTwo) rectangle ++ (12,\timelineheight);
	\draw[fill=timeLineBG, opacity=0.2] (0, \startYTwo) -- (0, \startYTwo + \timelineheight);
	\draw[fill=timeLineBG, opacity=0.2] (0.5, \startYTwo) -- (0.5, \startYTwo + \tlsmallheight);
	\draw[fill=timeLineBG, opacity=0.2] (1, \startYTwo) -- (1,  \startYTwo + \tlsmallheight);
	\draw[fill=timeLineBG, opacity=0.2] (1.5, \startYTwo) -- (1.5,  \startYTwo + \tlsmallheight);
	\draw[fill=timeLineBG, opacity=0.2] (2, \startYTwo) -- (2,  \startYTwo + \tlsmallheight);
	\draw[fill=timeLineBG, opacity=0.2] (2.5, \startYTwo) -- (2.5,  \startYTwo + \tlsmallheight);
	\draw[fill=timeLineBG, opacity=0.2] (3, \startYTwo) -- (3, \startYTwo + \timelineheight);
	\draw[fill=timeLineBG, opacity=0.2] (3.5, \startYTwo) -- (3.5, \startYTwo + \tlsmallheight);
	\draw[fill=timeLineBG, opacity=0.2] (4, \startYTwo) -- (4,  \startYTwo + \tlsmallheight);
	\draw[fill=timeLineBG, opacity=0.2] (4.5, \startYTwo) -- (4.5,  \startYTwo + \tlsmallheight);
	\draw[fill=timeLineBG, opacity=0.2] (5, \startYTwo) -- (5,  \startYTwo + \tlsmallheight);
	\draw[fill=timeLineBG, opacity=0.2] (5.5, \startYTwo) -- (5.5,  \startYTwo + \tlsmallheight);
	\draw[fill=timeLineBG, opacity=0.2] (6, \startYTwo) -- (6, \startYTwo + \timelineheight);
	\draw[fill=timeLineBG, opacity=0.2] (6.5, \startYTwo) -- (6.5, \startYTwo + \tlsmallheight);
	\draw[fill=timeLineBG, opacity=0.2] (7, \startYTwo) -- (7,  \startYTwo + \tlsmallheight);
	\draw[fill=timeLineBG, opacity=0.2] (7.5, \startYTwo) -- (7.5,  \startYTwo + \tlsmallheight);
	\draw[fill=timeLineBG, opacity=0.2] (8, \startYTwo) -- (8,  \startYTwo + \tlsmallheight);
	\draw[fill=timeLineBG, opacity=0.2] (8.5, \startYTwo) -- (8.5,  \startYTwo + \tlsmallheight);
	\draw[fill=timeLineBG, opacity=0.2] (9, \startYTwo) -- (9, \startYTwo + \timelineheight);
	\draw[fill=timeLineBG, opacity=0.2] (9.5, \startYTwo) -- (9.5, \startYTwo + \tlsmallheight);
	\draw[fill=timeLineBG, opacity=0.2] (10, \startYTwo) -- (10,  \startYTwo + \tlsmallheight);
	\draw[fill=timeLineBG, opacity=0.2] (10.5, \startYTwo) -- (10.5,  \startYTwo + \tlsmallheight);
	\draw[fill=timeLineBG, opacity=0.2] (11, \startYTwo) -- (11,  \startYTwo + \tlsmallheight);
	\draw[fill=timeLineBG, opacity=0.2] (11.5, \startYTwo) -- (11.5,  \startYTwo + \tlsmallheight);
	\draw[fill=timeLineBG, opacity=0.2] (12, \startYTwo) -- (12, \startYTwo + \timelineheight);
	% Year labels
	\node (Year2016) at (1.5, \startYTwo + \timelineheight) [above, color=gray] {2016};
	\node (Year2017) at (4.5, \startYTwo + \timelineheight) [above, color=gray] {2017};
	\node (Year2018) at (7.5, \startYTwo + \timelineheight) [above, color=gray] {2018};
	\node (Year2019) at (10.5, \startYTwo + \timelineheight) [above, color=gray] {2019};
	%
	% 2018
	\node (Rueegg18) at (6 + \leftMargin, \startYTwo + \bottomMargin) [above, color=gray] {\cite{Rueegg18}};
	\node (RueeggvH18a) at (7 + \leftMargin, \startYTwo + \elkheight + \bottomMargin) [above, color=gray] {\cite{RueeggvH18a}};
	\node (SchulzeWvH18a) at (7 + \leftMargin, \startYTwo + \elkheight + \bottomMargin + \fontheight + \padding) [above, color=gray] {\cite{SchulzeWvH18a}};
	\node (SchulzeHvH18a) at (8 + \leftMargin, \startYTwo + \bottomMargin) [above, color=gray] {\cite{SchulzeHvH18a}};
	% 2019
	\node (Schulze19) at (10.5 + \leftMargin, \startYTwo + \bottomMargin) [above, color=gray] {\cite{Schulze19}};
	% ELK triangles
	% 2016
	\fill[above, fill=gray] (1.5 + \etlmargin, \startYTwo + \etbmargin) node[anchor=north]{}
	-- (1.5 + \etlmargin + \etwidth, \startYTwo + \etheight + \etbmargin) node[anchor=north]{}
	-- (1.5 + \etlmargin + \etwidth + \etwidth, \startYTwo + \etbmargin) node[anchor=south]{}
	-- cycle;
	% 2017
	\fill[above, fill=gray] (3 + \etlmargin, \startYTwo + \etbmargin) node[anchor=north]{}
	-- (3 + \etlmargin + \etwidth, \startYTwo + \etheight + \etbmargin) node[anchor=north]{}
	-- (3 + \etlmargin + \etwidth + \etwidth, \startYTwo + \etbmargin) node[anchor=south]{}
	-- cycle;
	\fill[above, fill=gray] (4.5 + \etlmargin, \startYTwo + \etbmargin) node[anchor=north]{}
	-- (4.5 + \etlmargin + \etwidth, \startYTwo + \etheight + \etbmargin) node[anchor=north]{}
	-- (4.5 + \etlmargin + \etwidth + \etwidth, \startYTwo + \etbmargin) node[anchor=south]{}
	-- cycle;
	\fill[above, fill=gray] (5 + \etlmargin, \startYTwo + \etbmargin) node[anchor=north]{}
	-- (5 + \etlmargin + \etwidth, \startYTwo + \etheight + \etbmargin) node[anchor=north]{}
	-- (5 + \etlmargin + \etwidth + \etwidth, \startYTwo + \etbmargin) node[anchor=south]{}
	-- cycle;
	% 2018
	\fill[above, fill=gray] (7 + \etlmargin, \startYTwo + \etbmargin) node[anchor=north]{}
	-- (7 + \etlmargin + \etwidth, \startYTwo + \etheight + \etbmargin) node[anchor=north]{}
	-- (7 + \etlmargin + \etwidth + \etwidth, \startYTwo + \etbmargin) node[anchor=south]{}
	-- cycle;
	% 2019
	\fill[above, fill=gray] (9.5 + \etlmargin, \startYTwo + \etbmargin) node[anchor=north]{}
	-- (9.5 + \etlmargin + \etwidth, \startYTwo + \etheight + \etbmargin) node[anchor=north]{}
	-- (9.5 + \etlmargin + \etwidth + \etwidth, \startYTwo + \etbmargin) node[anchor=south]{}
	-- cycle;
	\fill[above, fill=gray] (11.5 + \etlmargin, \startYTwo + \etbmargin) node[anchor=north]{}
	-- (11.5 + \etlmargin + \etwidth, \startYTwo + \etheight + \etbmargin) node[anchor=north]{}
	-- (11.5 + \etlmargin + \etwidth + \etwidth, \startYTwo + \etbmargin) node[anchor=south]{}
	-- cycle;
	% ELK Labels
	% 2016
	\node (ELK010) at (1.75, \startYTwo + \elheight) [text centered, color=gray] {0.1.0};
	% 2017
	\node (ELK020) at (3.4, \startYTwo + \elheight) [text centered, color=gray] {0.2.0};
	\node (elkjs) at (4.6, \startYTwo + \elheight) [text centered, color=gray] {elkjs};
	\node (ELK030) at (5.4, \startYTwo + \elheight) [text centered, color=gray] {0.3.0};
	% 2018
	\node (ELK040) at (7.25, \startYTwo + \elheight) [text centered, color=gray] {0.4.0};
	% 2019
	\node (ELK050) at (9.75, \startYTwo + \elheight) [text centered, color=gray] {0.5.0};
	\node (ELK060) at (11.6, \startYTwo + \elheight) [text centered, color=gray] {0.6.0};
	% 2016
	\node (RueeggLPK+16) at (2 + \leftMargin, \startYTwo + \bottomMargin) [above, color=gray] {\cite{RueeggLPK+16}};
	\node (JabrayilovMMR+16) at (2 + \leftMargin, \startYTwo + \padding + \fontheight + \bottomMargin) [above, color=gray] {\cite{JabrayilovMMR+16}};
	\node (Schulze16) at (2 + \leftMargin, \startYTwo + \padding + \padding + \fontheight + \fontheight + \bottomMargin) [above, color=gray] {\cite{Schulze16}};
	\node (RueeggESvH16) at (2.5 + \leftMargin, \startYTwo + \bottomMargin) [above, color=gray] {\cite{RueeggESvH16}};
	\node (RueeggSSW+16) at (2.5 + \leftMargin, \startYTwo + \padding + \fontheight + \bottomMargin) [above, color=gray] {\cite{RueeggSSW+16}};
	\node (SchulzeLvH16) at (2.5 + \leftMargin, \startYTwo + \padding + \padding + \fontheight + \fontheight + \bottomMargin) [above, color=gray] {\cite{SchulzeLvH16}};
	\node (RueeggSGvH16b) at (1.5 + \leftMargin, \startYTwo + \elkheight + \bottomMargin) [above, color=gray] {\cite{RueeggSGvH16b}};
	% 2016
	% MT Schelten
	\draw[color=gold, line width=\thlinethick] (0.75, \startYTwo - \thHeighth) -- (0.75, \startYTwo + \thHeighth);\thHeighth
	\draw[color=gold, line width=\thlinethick] (0.75, \startYTwo) -- (0.75 + \monthWidth * 6, \startYTwo);\thHeighth
	\draw[color=gold, line width=\thlinethick] (0.75 + \monthWidth * 6, \startYTwo - \thHeighth) -- (0.75 + \monthWidth * 6, \startYTwo + \thHeighth);
	\node (MTSchelten) at (1.5, \startYTwo - \thlineheight - \thLabelHeight) [text centered, color=gold] {\scriptsize MT Schelten};
	% MT Sprung
	\draw[color=gold, line width=\thlinethick] (1, \startYTwo - \thHeighth - \thlineheight) -- (1, \startYTwo + \thHeighth - \thlineheight);\thHeighth
	\draw[color=gold, line width=\thlinethick] (1, \startYTwo - \thlineheight) -- (1 + \monthWidth * 6, \startYTwo - \thlineheight);\thHeighth
	\draw[color=gold, line width=\thlinethick] (1 + \monthWidth * 6, \startYTwo - \thHeighth - \thlineheight) -- (1 + \monthWidth * 6, \startYTwo + \thHeighth - \thlineheight);
	\node (MTSprung) at (2, \startYTwo - \thlineheight - \thLabelHeight - \thLabelHeight -0.1) [text centered, color=gold] {\scriptsize MT Sprung};
	% MT Cyruk first part
	\draw[color=gold, line width=\thlinethick] (2.75, \startYTwo - \thHeighth) -- (2.75, \startYTwo + \thHeighth);\thHeighth
	\draw[color=gold, line width=\thlinethick] (2.75, \startYTwo) -- (2.75 + \monthWidth, \startYTwo);\thHeighth
	% 2017
	% MT Cyruk
	\draw[color=gold, line width=\thlinethick] (3, \startYTwo) -- (3 + \monthWidth * 5, \startYTwo);\thHeighth
	\draw[color=gold, line width=\thlinethick] (3 + \monthWidth * 5, \startYTwo - \thHeighth) -- (3 + \monthWidth * 5, \startYTwo + \thHeighth);
	\node (MTCyruk) at (3.75, \startYTwo - \thLabelHeight) [text centered, color=gold] {\scriptsize MT Cyruk};
	% MT Jahn
	\draw[color=gold, line width=\thlinethick] (4.5, \startYTwo - \thHeighth) -- (4.5, \startYTwo + \thHeighth);\thHeighth
	\draw[color=gold, line width=\thlinethick] (4.5, \startYTwo) -- (4.5 + \monthWidth * 6, \startYTwo);\thHeighth
	\draw[color=gold, line width=\thlinethick] (4.5 + \monthWidth * 6, \startYTwo - \thHeighth) -- (4.5 + \monthWidth * 6, \startYTwo + \thHeighth);
	\node (MTJahn) at (5.25, \startYTwo - \thLabelHeight) [text centered, color=gold] {\scriptsize MT Jahn};
	% 2018
	% BT Lucas
	\draw[color=gold, line width=\thlinethick] (6.75, \startYTwo - \thHeighth) -- (6.75, \startYTwo + \thHeighth);\thHeighth
	\draw[color=gold, line width=\thlinethick] (6.75, \startYTwo) -- (6.75 + \monthWidth * 6, \startYTwo);\thHeighth
	\draw[color=gold, line width=\thlinethick] (6.75 + \monthWidth * 6, \startYTwo - \thHeighth) -- (6.75 + \monthWidth * 6, \startYTwo + \thHeighth);
	\node (BTLucas) at (7.5, \startYTwo - \thLabelHeight) [text centered, color=gold] {\scriptsize BT Lucas};
	% BT Weber/BT Borgfeld
	\draw[color=gold, line width=\thlinethick] (8.25, \startYTwo - \thHeighth) -- (8.25, \startYTwo + \thHeighth);\thHeighth
	\draw[color=gold, line width=\thlinethick] (8.25, \startYTwo) -- (8.25 + \monthWidth * 6, \startYTwo);\thHeighth
	\draw[color=gold, line width=\thlinethick] (8.25 + \monthWidth * 6, \startYTwo - \thHeighth) -- (8.25 + \monthWidth * 6, \startYTwo + \thHeighth);
	\node (BTWeber) at (9, \startYTwo - \thLabelHeight) [text centered, color=gold] {\scriptsize BT Weber};
	\node (BTBorgfeld) at (9, \startYTwo - \thLabelHeight - \thlineheight -0.1) [text centered, color=gold] {\scriptsize BT Borgfeld};
	% 2019
	% BT Schönberner/BT Petzold
	\draw[color=gold, line width=\thlinethick] (9.75, \startYTwo - \thHeighth) -- (9.75, \startYTwo + \thHeighth);\thHeighth
	\draw[color=gold, line width=\thlinethick] (9.75, \startYTwo) -- (9.75 + \monthWidth * 6, \startYTwo);\thHeighth
	\draw[color=gold, line width=\thlinethick] (9.75 + \monthWidth * 6, \startYTwo - \thHeighth) -- (9.75 + \monthWidth * 6, \startYTwo + \thHeighth);
	\node (BTSchönberner) at (10.7, \startYTwo - \thLabelHeight) [text centered, color=gold] {\scriptsize BT Schönberner};
	\node (BTPetzold) at (10.7, \startYTwo - \thLabelHeight - \thlineheight -0.1) [text centered, color=gold] {\scriptsize BT Petzold};
	\fill[fill=timeLineBG, fill opacity=0.1] (0, \startYThree) rectangle ++ (12,\timelineheight);
	\draw[fill=timeLineBG, opacity=0.2] (0, \startYThree) -- (0, \startYThree + \timelineheight);
	\draw[fill=timeLineBG, opacity=0.2] (0.5, \startYThree) -- (0.5, \startYThree + \tlsmallheight);
	\draw[fill=timeLineBG, opacity=0.2] (1, \startYThree) -- (1,  \startYThree + \tlsmallheight);
	\draw[fill=timeLineBG, opacity=0.2] (1.5, \startYThree) -- (1.5,  \startYThree + \tlsmallheight);
	\draw[fill=timeLineBG, opacity=0.2] (2, \startYThree) -- (2,  \startYThree + \tlsmallheight);
	\draw[fill=timeLineBG, opacity=0.2] (2.5, \startYThree) -- (2.5,  \startYThree + \tlsmallheight);
	\draw[fill=timeLineBG, opacity=0.2] (3, \startYThree) -- (3, \startYThree + \timelineheight);
	\draw[fill=timeLineBG, opacity=0.2] (3.5, \startYThree) -- (3.5, \startYThree + \tlsmallheight);
	\draw[fill=timeLineBG, opacity=0.2] (4, \startYThree) -- (4,  \startYThree + \tlsmallheight);
	\draw[fill=timeLineBG, opacity=0.2] (4.5, \startYThree) -- (4.5,  \startYThree + \tlsmallheight);
	\draw[fill=timeLineBG, opacity=0.2] (5, \startYThree) -- (5,  \startYThree + \tlsmallheight);
	\draw[fill=timeLineBG, opacity=0.2] (5.5, \startYThree) -- (5.5,  \startYThree + \tlsmallheight);
	\draw[fill=timeLineBG, opacity=0.2] (6, \startYThree) -- (6, \startYThree + \timelineheight);
	\draw[fill=timeLineBG, opacity=0.2] (6.5, \startYThree) -- (6.5, \startYThree + \tlsmallheight);
	\draw[fill=timeLineBG, opacity=0.2] (7, \startYThree) -- (7,  \startYThree + \tlsmallheight);
	\draw[fill=timeLineBG, opacity=0.2] (7.5, \startYThree) -- (7.5,  \startYThree + \tlsmallheight);
	\draw[fill=timeLineBG, opacity=0.2] (8, \startYThree) -- (8,  \startYThree + \tlsmallheight);
	\draw[fill=timeLineBG, opacity=0.2] (8.5, \startYThree) -- (8.5,  \startYThree + \tlsmallheight);
	\draw[fill=timeLineBG, opacity=0.2] (9, \startYThree) -- (9, \startYThree + \timelineheight);
	\draw[fill=timeLineBG, opacity=0.2] (9.5, \startYThree) -- (9.5, \startYThree + \tlsmallheight);
	\draw[fill=timeLineBG, opacity=0.2] (10, \startYThree) -- (10,  \startYThree + \tlsmallheight);
	\draw[fill=timeLineBG, opacity=0.2] (10.5, \startYThree) -- (10.5,  \startYThree + \tlsmallheight);
	\draw[fill=timeLineBG, opacity=0.2] (11, \startYThree) -- (11,  \startYThree + \tlsmallheight);
	\draw[fill=timeLineBG, opacity=0.2] (11.5, \startYThree) -- (11.5,  \startYThree + \tlsmallheight);
	\draw[fill=timeLineBG, opacity=0.2] (12, \startYThree) -- (12, \startYThree + \timelineheight);
	% Year labels
	\node (Year2012) at (1.5, \startYThree + \timelineheight) [above, color=gray] {2012};
	\node (Year2013) at (4.5, \startYThree + \timelineheight) [above, color=gray] {2013};
	\node (Year2014) at (7.5, \startYThree + \timelineheight) [above, color=gray] {2014};
	\node (Year2015) at (10.5, \startYThree + \timelineheight) [above, color=gray] {2015};
	% Pubs
	% 2012
	\node (KlauskeSS+12) at (1.5 + \leftMargin, \startYThree + \elkheight + \bottomMargin) [above, color=gray] {\cite{KlauskeSS+12}};
	% 2014
	\node (SchulzeSSvH14) at (6.5 + \leftMargin, \startYThree + \bottomMargin) [above, color=gray] {\cite{SchulzeSSvH14}};
	\node (SpoenemannDvH14b) at (7.5 + \leftMargin, \startYThree + \bottomMargin) [above, color=gray] {\cite{SpoenemannDvH14b}};
	\node (GutwengervHM+14) at (7.5 + \leftMargin, \startYThree + \padding + \fontheight + \bottomMargin) [above, color=gray] {\cite{GutwengervHM+14}};
	\node (SpoenemannSRvH14) at (7.5 + \leftMargin, \startYThree + \padding + \padding + \fontheight + \fontheight + \bottomMargin) [above, color=gray] {\cite{SpoenemannSRvH14}};
	% 2015
	\node (Spoenemann15) at (9 + \leftMargin, \startYThree + \bottomMargin) [above, color=gray] {\cite{Spoenemann15}};
	\node (RueeggSCvH15) at (11 + \leftMargin, \startYThree + \bottomMargin) [above, color=gray] {\cite{RueeggSCvH15}};
	% ELK triangles
	% 2012
	\fill[above, fill=gray] (1.5 + \etlmargin, \startYThree + \etbmargin) node[anchor=north]{}
	-- (1.5 + \etlmargin + \etwidth, \startYThree + \etheight + \etbmargin) node[anchor=north]{}
	-- (1.5 + \etlmargin + \etwidth + \etwidth, \startYThree + \etbmargin) node[anchor=south]{}
	-- cycle;
	% 2015
	\fill[above, fill=gray] (10.5 + \etlmargin, \startYThree + \etbmargin) node[anchor=north]{}
	-- (10.5 + \etlmargin + \etwidth, \startYThree + \etheight + \etbmargin) node[anchor=north]{}
	-- (10.5 + \etlmargin + \etwidth + \etwidth, \startYThree + \etbmargin) node[anchor=south]{}
	-- cycle;
	% ELK Labels
	% 2012
	\node (KLay layered) at (1.75, \startYThree + \elheight) [text centered, color=gray] {KLay Layered};
	% 2015
	\node (ELKCreation) at (10.75, \startYThree + \elheight) [text centered, color=gray] {ELK Creation};
	% 2012
	% MT Carstens
	\draw[color=gold, line width=\thlinethick] (0.75, \startYThree - \thHeighth) -- (0.75, \startYThree + \thHeighth);\thHeighth
	\draw[color=gold, line width=\thlinethick] (0.75, \startYThree) -- (0.75 + \monthWidth * 6, \startYThree);\thHeighth
	\draw[color=gold, line width=\thlinethick] (0.75 + \monthWidth * 6, \startYThree - \thHeighth) -- (0.75 + \monthWidth * 6, \startYThree + \thHeighth);
	\node (MTCartens) at (1.5, \startYThree - \thlineheight - \thLabelHeight -0.1) [text centered, color=gold] {\scriptsize MT Cartens};
	% 2014
	% DT Furhmann second part
	\draw[color=gold, line width=\thlinethick] (0, \startYThree) -- (0 + \monthWidth * 2, \startYThree);\thHeighth
	\draw[color=gold, line width=\thlinethick] (0 + \monthWidth * 2, \startYThree - \thHeighth) -- (0 + \monthWidth * 2, \startYThree + \thHeighth);
	\node (DTFuhrmann) at (1, \startYThree - \thLabelHeight) [color=gold] {\scriptsize DT Fuhrmann};
	% DT Toepffer
	\draw[color=gold, line width=\thlinethick] (7.25, \startYThree - \thHeighth) -- (7.25, \startYThree + \thHeighth);\thHeighth
	\draw[color=gold, line width=\thlinethick] (7.25, \startYThree) -- (7.25 + \monthWidth * 6, \startYThree);\thHeighth
	\draw[color=gold, line width=\thlinethick] (7.25 + \monthWidth * 6, \startYThree - \thHeighth) -- (7.25 + \monthWidth * 6, \startYThree + \thHeighth);
	\node (DTToepffer) at (7.4, \startYThree - \thLabelHeight) [text centered, color=gold] {\scriptsize DT Toepffer};
	% BT Schelten
	\draw[color=gold, line width=\thlinethick] (8.25, \startYThree - \thHeighth - \thlineheight) -- (8.25, \startYThree + \thHeighth - \thlineheight);\thHeighth
	\draw[color=gold, line width=\thlinethick] (8.25, \startYThree - \thlineheight) -- (8.25 + \monthWidth * 6, \startYThree - \thlineheight);\thHeighth
	\draw[color=gold, line width=\thlinethick] (8.25 + \monthWidth * 6, \startYThree - \thHeighth - \thlineheight) -- (8.25 + \monthWidth * 6, \startYThree + \thHeighth - \thlineheight);
	\node (BTSchelten) at (9, \startYThree - \thlineheight - \thLabelHeight) [text centered, color=gold] {\scriptsize BT Schelten};
	% 2015
	% BT Lasch
	\draw[color=gold, line width=\thlinethick] (9.75, \startYThree - \thHeighth) -- (9.75, \startYThree + \thHeighth);\thHeighth
	\draw[color=gold, line width=\thlinethick] (9.75, \startYThree) -- (9.75 + \monthWidth * 6, \startYThree);\thHeighth
	\draw[color=gold, line width=\thlinethick] (9.75 + \monthWidth * 6, \startYThree - \thHeighth) -- (9.75 + \monthWidth * 6, \startYThree + \thHeighth);
	\node (BTLasch) at (10.5, \startYThree - \thLabelHeight) [text centered, color=gold] {\scriptsize BT Lasch};
	\fill[fill=timeLineBG, fill opacity=0.1] (0, \startYFour) rectangle ++ (12,\timelineheight);
	\draw[fill=timeLineBG, opacity=0.2] (0, \startYFour) -- (0, \startYFour + \timelineheight);
	\draw[fill=timeLineBG, opacity=0.2] (0.5, \startYFour) -- (0.5, \startYFour + \tlsmallheight);
	\draw[fill=timeLineBG, opacity=0.2] (1, \startYFour) -- (1,  \startYFour + \tlsmallheight);
	\draw[fill=timeLineBG, opacity=0.2] (1.5, \startYFour) -- (1.5,  \startYFour + \tlsmallheight);
	\draw[fill=timeLineBG, opacity=0.2] (2, \startYFour) -- (2,  \startYFour + \tlsmallheight);
	\draw[fill=timeLineBG, opacity=0.2] (2.5, \startYFour) -- (2.5,  \startYFour + \tlsmallheight);
	\draw[fill=timeLineBG, opacity=0.2] (3, \startYFour) -- (3, \startYFour + \timelineheight);
	\draw[fill=timeLineBG, opacity=0.2] (3.5, \startYFour) -- (3.5, \startYFour + \tlsmallheight);
	\draw[fill=timeLineBG, opacity=0.2] (4, \startYFour) -- (4,  \startYFour + \tlsmallheight);
	\draw[fill=timeLineBG, opacity=0.2] (4.5, \startYFour) -- (4.5,  \startYFour + \tlsmallheight);
	\draw[fill=timeLineBG, opacity=0.2] (5, \startYFour) -- (5,  \startYFour + \tlsmallheight);
	\draw[fill=timeLineBG, opacity=0.2] (5.5, \startYFour) -- (5.5,  \startYFour + \tlsmallheight);
	\draw[fill=timeLineBG, opacity=0.2] (6, \startYFour) -- (6, \startYFour + \timelineheight);
	\draw[fill=timeLineBG, opacity=0.2] (6.5, \startYFour) -- (6.5, \startYFour + \tlsmallheight);
	\draw[fill=timeLineBG, opacity=0.2] (7, \startYFour) -- (7,  \startYFour + \tlsmallheight);
	\draw[fill=timeLineBG, opacity=0.2] (7.5, \startYFour) -- (7.5,  \startYFour + \tlsmallheight);
	\draw[fill=timeLineBG, opacity=0.2] (8, \startYFour) -- (8,  \startYFour + \tlsmallheight);
	\draw[fill=timeLineBG, opacity=0.2] (8.5, \startYFour) -- (8.5,  \startYFour + \tlsmallheight);
	\draw[fill=timeLineBG, opacity=0.2] (9, \startYFour) -- (9, \startYFour + \timelineheight);
	\draw[fill=timeLineBG, opacity=0.2] (9.5, \startYFour) -- (9.5, \startYFour + \tlsmallheight);
	\draw[fill=timeLineBG, opacity=0.2] (10, \startYFour) -- (10,  \startYFour + \tlsmallheight);
	\draw[fill=timeLineBG, opacity=0.2] (10.5, \startYFour) -- (10.5,  \startYFour + \tlsmallheight);
	\draw[fill=timeLineBG, opacity=0.2] (11, \startYFour) -- (11,  \startYFour + \tlsmallheight);
	\draw[fill=timeLineBG, opacity=0.2] (11.5, \startYFour) -- (11.5,  \startYFour + \tlsmallheight);
	\draw[fill=timeLineBG, opacity=0.2] (12, \startYFour) -- (12, \startYFour + \timelineheight);
	% Year labels
	\node (Year2008) at (1.5, \startYFour + \timelineheight) [above, color=gray] {2008};
	\node (Year2009) at (4.5, \startYFour + \timelineheight) [above, color=gray] {2009};
	\node (Year2010) at (7.5, \startYFour + \timelineheight) [above, color=gray] {2010};
	\node (Year2011) at (10.5, \startYFour + \timelineheight) [above, color=gray] {2011};
	% Pubs
	% 2009
	\node (SpoenemannFvH09) at (3 + \leftMargin, \startYFour + \bottomMargin) [above, color=gray] {\cite{SpoenemannFvH09}};
	\node (SpoenemannFvH+09) at (5 + \leftMargin, \startYFour + \elkheight + \bottomMargin) [above, color=gray] {\cite{SpoenemannFvH+09}};
	% ELK triangles
	% 2009
	\fill[above, fill=gray] (5 + \etlmargin, \startYFour + \etbmargin) node[anchor=north]{}
	-- (5 + \etlmargin + \etwidth, \startYFour + \etheight + \etbmargin) node[anchor=north]{}
	-- (5 + \etlmargin + \etwidth + \etwidth, \startYFour + \etbmargin) node[anchor=south]{}
	-- cycle;
	% ELK Labels
	% 2009
	\node (KLoDD) at (5.25, \startYFour + \elheight) [text centered, color=gray] {KLoDD};
	% Theses Lines
	% 2009
	% DT Spoenemann
	\draw[color=gold, line width=\thlinethick] (2.25, \startYFour - \thHeighth) -- (2.25, \startYFour + \thHeighth);\thHeighth
	\draw[color=gold, line width=\thlinethick] (2.25, \startYFour) -- (2.25 + \monthWidth * 6, \startYFour);\thHeighth
	\draw[color=gold, line width=\thlinethick] (2.25 + \monthWidth * 6, \startYFour - \thHeighth) -- (2.25 + \monthWidth * 6, \startYFour + \thHeighth);
	\node (DTSpoenemann) at (3, \startYFour - \thLabelHeight) [text centered, color=gold]
	{\scriptsize DT Spönemann};
	% 2010
	% BT Rieß BT Döhring
	\draw[color=gold, line width=\thlinethick] (6.75, \startYFour - \thHeighth) -- (6.75, \startYFour + \thHeighth);\thHeighth
	\draw[color=gold, line width=\thlinethick] (6.75, \startYFour) -- (6.75 + \monthWidth * 6, \startYFour);\thHeighth
	\draw[color=gold, line width=\thlinethick] (6.75 + \monthWidth * 6, \startYFour - \thHeighth) -- (6.75 + \monthWidth * 6, \startYFour + \thHeighth);
	\node (BTRiess) at (7.5, \startYFour - \thLabelHeight) [text centered, color=gold] {\scriptsize BT Rieß};
	\node (BTDoehring) at (7.5, \startYFour - \thlineheight - \thLabelHeight - 0.1) [text centered, color=gold] {\scriptsize BT Döhring};
	% 2011
	% DT Schulze
	\draw[color=gold, line width=\thlinethick] (9.25, \startYFour - \thHeighth) -- (9.25, \startYFour + \thHeighth);\thHeighth
	\draw[color=gold, line width=\thlinethick] (9.25, \startYFour) -- (9.25 + \monthWidth * 6, \startYFour);\thHeighth
	\draw[color=gold, line width=\thlinethick] (9.25 + \monthWidth * 6, \startYFour - \thHeighth) -- (9.25 + \monthWidth * 6, \startYFour + \thHeighth);
	\node (DTSchulze) at (10, \startYFour - \thlineheight - \thLabelHeight) [text centered, color=gold] {\scriptsize DT Schulze};
	% DT Wersig
	\draw[color=gold, line width=\thlinethick] (9.75, \startYFour - \thHeighth - \thlineheight) -- (9.75, \startYFour + \thHeighth - \thlineheight);\thHeighth
	\draw[color=gold, line width=\thlinethick] (9.75, \startYFour - \thlineheight) -- (9.75 + \monthWidth * 6, \startYFour - \thlineheight);\thHeighth
	\draw[color=gold, line width=\thlinethick] (9.75 + \monthWidth * 6, \startYFour - \thHeighth - \thlineheight) -- (9.75 + \monthWidth * 6, \startYFour + \thHeighth - \thlineheight);
	\node (DTWersig) at (10.5, \startYFour - \thlineheight - \thLabelHeight - \thLabelHeight -0.1) [text centered, color=gold] {\scriptsize DT Wersig};
	% DT Fuhrmann first part
	\draw[color=gold, line width=\thlinethick] (11, \startYFour - \thHeighth) -- (11, \startYFour + \thHeighth);\thHeighth
	\draw[color=gold, line width=\thlinethick] (11, \startYFour) -- (11 + \monthWidth * 4, \startYFour);\thHeighth
}%
\end{tikzpicture}
	\caption{The \ac{elk} development timeline. Selected diploma theses (DT), bachelor theses (BT), and master theses (MT) are marked in green, software releases are marked with a gray triangle, and relevant publications are cited.}
	\label{fig:timeline}
\end{figure}
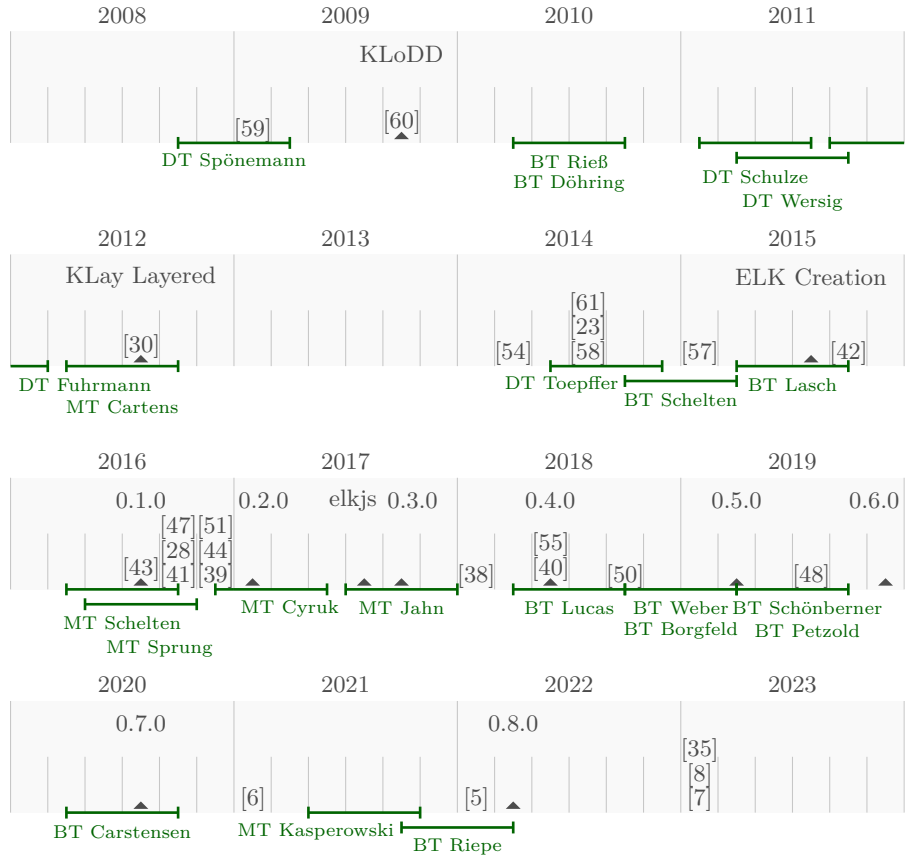
Years of work and several publications formed \ac{elk} as it is presented today:
An open-source layout library for layered layouts, rectangle packing, tree drawing, force and stress layouts, radial layouts, and packing of disconnected components,
which transpiles into the popular (see \autoref{fig:npm-trends}) npm-package elkjs\footnote{\url{https://github.com/kieler/elkjs}},
and includes bridges to Graphviz \cite{GansnerN00} and libavoid \cite{WybrowMS10}.

% \rsdo{Add refactoring point in the history of ELK. (no, not clear enough, only mention later)
% Note that this it is not expected for researchers to put that much effort into software engineering.
% Very few people work on ELK and but not full-time and they change after a few years.
% New people are hard to get.
% The well maintained infrastructure allows 
% }

In the following, we will continue with related work.
\autoref{sec:application} presents the users of \ac{elk} and elkjs.
\acposs{elk} layout algorithms are presented in \autoref{sec:algorithms} with special focus on \ac{elk} Layered.
We contribute an in depth discussion of \acposs{elk} architecture, the algorithm architecture of the \ac{elk} Layered algorithm, and lessons learned from developing and maintaining a large open-source layout algorithm library in an academic context in \autoref{sec:implementation}.
\autoref{sec:examples} presents real world examples layouted with \ac{elk}.
\autoref{sec:conclusion} concludes the paper and presents future work on \ac{elk}.

\subsubsection*{Related Work}
\label{sec:related-work}

There exist several layout libraries, graphical editors, or algorithms that utilize automatic layout, as presented by Jünger and Mutzel~\cite{JuengerM03}.

Dunnart \cite{DwyerMW09b} is a constraint based diagram editor, which utilizes constraints to create desirable layouts using libavoid \cite{WybrowMS10}, OGDF \cite{ChimaniG+07}, or cola \cite{DwyerK05}.

Cytoscape\cite{FranzLHD+16} is a layout library originally used for bio-informatics, which provides several algorithms for various use cases including webcola\footnote{\url{https://ialab.it.monash.edu/webcola/}}, dagre, and elkjs.
Similar to \ac{elk}, this allows using the same interface for various use cases.

The yWorks GmbH and their commercial yFiles tool \cite{WieseEK01} provide layout algorithms and visualizations.
However, it is not open-source and the free yEd tool does not expose an API only for layout.

Zink \etal \cite{ZinkWBW22} support layered layouts with port constraint groups.
Instead of just fixed positions or orderings, they allow grouping ports, which again may contain additional port groups that can have distinct layout constraints.
In terms of ports constraints this Sugiyama algorithm is superior.
However, it is especially developed for layout of hardware components and lacks \acposs{elk} configurability.

\section{ELK Availability, Usage and Algorithms}
\label{sec:application}

\ac{elk} mainly serves as a sandbox for developing new algorithms or layout strategies, which are often used in the context of \ac {mde} to create data-flow or control-flow diagrams \cite{vonHanxledenLF+22}.
Most publications are based on the layered approach \cite{SpoenemannFvH+09,
	FuhrmannvH10b,
	vHanxleden11,
	KlauskeSS+12,
	GutwengervHM+14,
	SchulzeSvH14,
	SpoenemannDvH14b,
	SpoenemannSRvH14,
	RueeggSCvH15,
	JabrayilovMMR+16,
	RueeggSGvH16b,
	RueeggESvH16,
	RueeggSSW+16,
	RueeggvH18a,
	DomroesvH22,
	DomroesRvH23
}.
The only other algorithm with a peer reviewed paper is \ac{elk} Rectangle Packing \cite{DomroesLvHJ21, DomroesLvHJ23}, which is used for the layout of SCCharts \cite{vonHanxledenDM+14}, as seen in \autoref{fig:motor}.

\subsubsection{Availability}
\label{sec:software}

\ac{elk} is available on GitHub\footnote{\url{https://github.com/eclipse/elk}} and is released via maven central under EPL-2.0, which is a weak copyleft license (except \ac{elk} libavoid, which is available under LGPL, a strong copyleft license).
The JavaScript library elkjs is also available on GitHub and published via npm\footnote{\url{https://www.npmjs.com/package/elkjs}}.
Further information regarding \ac{elk} can be found on its Eclipse Project website\footnote{\url{https://projects.eclipse.org/projects/modeling.elk}}, the Eclipse \ac{elk} website\footnote{\url{https://www.eclipse.org/elk/}}, and the elklive website\footnote{\url{https://rtsys.informatik.uni-kiel.de/elklive/index.html}}, which hosts a simple \ac{elk} graph editor that allows users to switch between different \ac{elk} versions, can convert between different \ac{elk} graph formats, and houses examples graphs.

\subsubsection{Usage}

Since \ac{elk} is open-source and added to several other libraries and tools, we can only partly trace its usage on the basis of questions, pull-request, and issues.
Notable visualization libraries that use \ac{elk} or elkjs are KLighD \cite{SchneiderSvH13},
	Eclipse Sprotty\footnote{\url{https://sprotty.org}}, Terrastruct\footnote{\url{https://terrastruct.com/}},
	Cytoscape,
	Mermaid\footnote{\url{https://mermaid.js.org/}}, and reaflow\footnote{\url{https://github.com/reaviz/reaflow}}.
The integration of elkjs into other tools might be the reason for its increasing weekly downloads since the Cytoscape and elkjs downloads seem to correlate (see \autoref{fig:npm-trends}).
These libraries make \ac{elk} available in tools such as
	Eclipse 4diac\footnote{\url{https://www.eclipse.org/4diac/}},
	Eclipse Sirius\footnote{\url{https://www.eclipse.org/sirius/}},
	Eclipse Papyrus\footnote{\url{https://www.eclipse.org/papyrus/}},
	plantUML\footnote{\url{https://plantuml.com/}},
	the GLSP\footnote{\url{https://www.eclipse.org/glsp/}},
	bigER \cite{GlaserB21},
	ExplorViz \cite{FittkauKH17},
	the \ac{kieler} \cite{vonHanxledenDM+14},
	Lingua Franca \cite{LohstrohMBL21},
	Capella Diagrams by Digitale Schiene Deutschland\footnote{\url{https://github.com/DSD-DBS/capellambse-context-diagrams}},
	Ptolemy II \cite{Ptolemaeus14},
	ETAS EHandbook \cite{FreyvHK+14},
	and probably many more.
Since \ac{elk} is published as an Eclipse project, industrial users such as ETAS, which use \ac{elk} nearly since it creation, can safely rely on \ac{elk} without fearing issues regarding intellectual property, since every \ac{elk} committer has to sign the Eclipse Contributor Agreement\footnote{\url{https://www.eclipse.org/legal/ECA.php}}.
This enables \ac{elk} users that reach from model-driven engineers to doctors that look at medication plans layouted with \ac{elk} to use the library.

\subsubsection{Algorithms}
All these tools mainly use layered layouts (\ac{elk} Layered).
However, \ac{elk} also supports rectangle packing (\ac{elk} Box, \ac{elk} Rectangle Packing), tree drawing (\ac{elk} Mr. Tree), force and stress layout (\ac{elk} Force, \ac{elk} Stress), disconnected component packing (\ac{elk} DisCo), radial layout (\ac{elk} Radial), and topology aware compaction and overlap removal (\ac{elk} SpOre Compaction/Overlap Removal).
Moreover, it integrates the Graphviz layout algorithms Circo, Dot, FDP, Neato, and Twopi and a bridge to libavoid \cite{WybrowMS10}, which does standalone edge routing.
Previous versions of \ac{elk} also included a bridge to OGDF \cite{ChimaniG+07}.

Each layout algorithm might be fairly complex.
\Eg \ac{elk} Layered supports in its current state 140 layout options to further configure the Sugiyama algorithm to add spacing and configurations for edges, nodes, comments \cite{SchulzevH14}, labels \cite{SchulzePvH16a}, disconnected components \cite{Schulze19}, and ports \cite{SchulzeSvH14}.
Moreover, these layout options can configure and enable different layout strategies such as model order \cite{DomroesRvH23}, node \cite{PetzoldDSvH23} and port constraints \cite{SchulzeSvH14}, node size constraints, compaction \cite{RueeggSCvH15}, edge wrapping \cite{RueeggvH18a}, edge and node label placement \cite{SchulzeLvH16}, self-loop arrangement, top-down layout, and layout direction.

Additionally to the works mentioned in \autoref{fig:timeline}, \ac{elk} implements the following strategies based on works of several researchers:
	stress-majorization layout based on Gansner \etal \cite{GansnerKN05},
	Fruchtermann and Reingold force layout \cite{FruchtermanR91},
	Eades force layout \cite{Eades84}, and
	tree node placement based on Walker \cite{Walker90}.
Furthermore, the following algorithms are adopted for \ac{elk} Layered:
	a greedy cycle breaking algorithm inspired by Eades \etal and di Battista \etal \cite{EadesLS93, DiBattistaETT99},
	a depth-first cycle breaker by Gansner \etal \cite{GansnerKNV93},
	a node promotion algorithm used as layer assignment post-processing, a minimal width layerer, a stretch width layerer, and a longest path layerer by Nikolov \etal \cite{NikolovTB05, NikolovT06},
	Coffman-Graham layer assignment \cite{CoffmanG72},
	network-simplex layer assignment and node placement by Gansner \etal \cite{GansnerKNV93},
	Forster constraint resolving during barycenter crossing minimization \cite{Forster05},
	Sugiyama style crossing minimization\cite{SugiyamaTT81},
	Brandes and Köpf node placement \cite{BrandesK02},
	linear segment node placement based on Sander \cite{Sander96a},
	hyper-edge cycle detection based work of Eades \etal \cite{EadesLS93},
	orthogonal edge routing based on Sander and Di Battista \etal \cite{Sander96a, DiBattistaETT99},
	spline edge routing based on the approach of Sederberg \cite{Sederberg05}, and
	a scan-line algorithm based on Lengauer\cite{Lengauer90}.

\section{ELK Features}
\label{sec:algorithms}

In the following, we present how \ac{elk} Layered deals with compound graphs and model order as an example for two \ac{elk} features.

\subsubsection{Compound Graphs and Hierarchical Ports}
\begin{figure}
	\hspace*{\fill}
	\subfloat[]{
		\centering
		\includegraphics[scale=0.69]{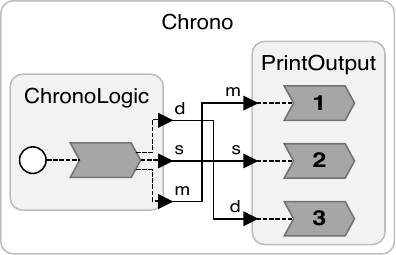}
		\label{fig:lf-chrono-ono}
	}
	\hfill
	\subfloat[]{
		\centering
		\includegraphics[scale=0.69]{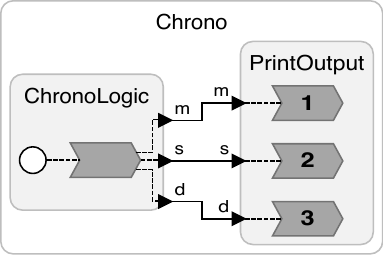}
		\label{fig:lf-chrono-oyes}
	}
	\hspace*{\fill}
	\caption{An abbreviated \textsff{Chrono} Lingua Franca \cite{LohstrohMBL21} program by Schulz-Rosengarten and von Hanxleden layouted bottom-up with \ac{elk}.
	In (a) the hierarchical edges may create preventable edge crossings, which can be prevented by considering the edges across hierarchy levels, as seen in (b).}
	\label{fig:lf-chrono}
\end{figure}

%...
Compound graphs or hierarchical graphs are laid out bottom-up by default.
The inner graph of a compound node determines the position of hierarchical ports, which cannot be changed by the parent.
In \autoref{fig:lf-chrono-ono} the order of the inner graph inside \textsff{ChronoLogic}, specifically the order of the outputs of the inner reaction (gray arrow without a number), determines the order of the hierarchical ports \textsff{d}, \textsff{s}, and \textsff{m} on the \textsff{ChronoLogic} node.
At the same time, the order of the reactions 1 - 3 (numbered gray arrows) inside \textsff{PrintOutput} determines the order of the hierarchical input ports \textsff{m}, \textsff{s}, and \textsff{d} of \textsff{PrintOutput}.
If these ordering decisions are made independently, edge crossings, as the one in \autoref{fig:lf-chrono-ono}, cannot be prevented.
\ac{elk} Layered can solve this problem by considering the hierarchical edges and their destination in the child nodes if desired.
This increases the complexity of the layout step and is invasive in the algorithm structure.
Therefore, many strategies such as different layout directions in different hierarchies and all strategies that influence ordering do not work together with this option.

\subsubsection{Model Order}
% ...
Model order represents the order of the input model,
which can often be used for layout decisions,
since \enquote{the input data can be expected to determine the choice \dots} (see Jünger and Mutzel  \cite{JuengerM03}, p. 31).
Domrös \etal describe how \ac{elk} Layered utilizes model order as a tie-breaker or a constraint not only during cycle breaking but also during layer assignment and crossing minimization to capture the user intention.
This approach tends to create better layouts that can be influenced more directly by users and translates secondary notation from a (textual) input model into the layout.
Furthermore, it solves common problems created by the bottom-up layout approach for compound graphs, as seen in \autoref{fig:lf-chrono}.
If one assumes that the user will always order ports, outputs, and inputs such that \textsff{m}, \textsff{s}, and \textsff{d} are consistently ordered, model order prevents the problem in \autoref{fig:lf-chrono-ono} form occurring.
Model order crossing minimization will create the crossing-free drawing depicted in \autoref{fig:lf-chrono-oyes} without using a complex hierarchy-aware layout algorithm.

\section{Implementation}
\label{sec:implementation}

\ac{elk} consists of 146000 lines of handwritten Java code (nearly 60000 lines alone for \ac{elk} Layered), 3500 lines of algorithm metadata, which generate the layout options and algorithm documentations, and five Xtext \acp{dsl}.
The Google Web Toolkit transpiles \ac{elk}-native algorithms into the JavaScript package elkjs to make \ac{elk} usable in web-applications.
However, elkjs is not as performant for larger graphs as the original \ac{elk}\footnote{\url{https://github.com/kieler/elk-speed}}.

\begin{figure}
	\centering
	\includegraphics[scale=0.45]{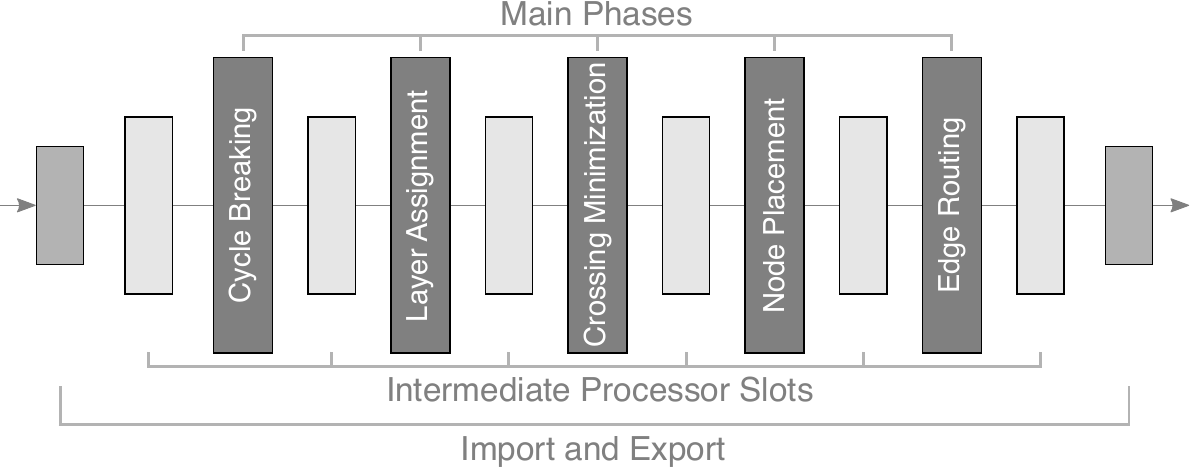}
	\caption{The five phases (dark gray) of \ac{elk} Layered with intermediate processor slots (light gray) before, between, and after each phase with an enclosing import and export step (gray) \cite{SchulzeSvH14}.}
	\label{fig:processor-slots}
\end{figure}

\subsection{Algorithm Architecture}

\ac{elk} provides a phase and processor infrastructure with import and export steps, as seen in \autoref{fig:processor-slots}.
The structuring of \ac{elk} Layered into five instead of the traditional three phases makes it easier to interchange different layout strategies.
The import step transforms a general \ac{elk} graph into a layered graph and transforms the layout problem into a left-to-right layout.
The exporter reverses these changes after the layout.
This allows algorithm-specific data-structures and abstracts from problems such as layout direction or implicit ports.
Additionally, pre- and post-processing may be done in the six intermediate processing slots.

\begin{figure}
	\centering
	\includegraphics[scale=0.45]{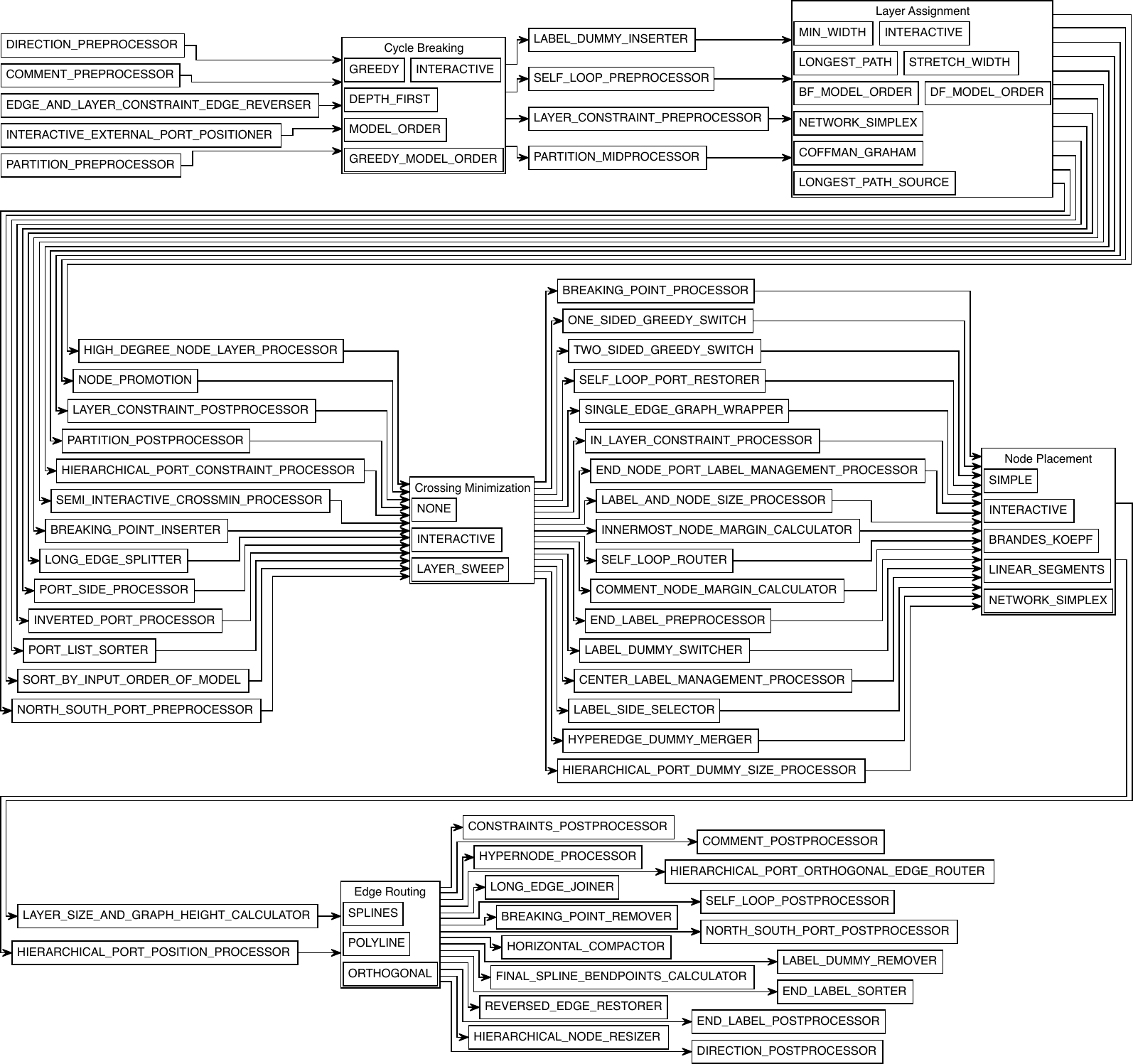}
	\caption{All \ac{elk} Layered processors, ordered by their execution order, and layout strategies, layouted with \ac{elk} Layered and styled with KLighD \cite{SchneiderSvH13}.
	}
	\label{fig:processor-all}
\end{figure}

Before, between, and after each of the phases, each of the 57 intermediate processors (see \autoref{fig:processor-all}) can be executed  with the limitation that they need to have a static execution order \cite{SchulzeSvH14}.
The use of intermediate processors might slightly increase the running time of the algorithm. However, these processors make the algorithm much more maintainable by avoiding duplicate code and allowing to adopt algorithms for different sub-problems.
This is done programmatically to keep the phase-processor dependencies maintainable.
\Eg, the {\scriptsize \textsff{GREEDY}} cycle breaking strategy can define that it needs the {\scriptsize \textsff{REVERSED\_EDGE\_RESTORER}} processor, which re-reverses the reversed edges, to be executed after phase five.

Since \ac{elk} has its roots in the Eclipse \ac{rca} \ac{kieler}, it additionally contributes debugging and logging views for Eclipse \ac{ide}.
Furthermore, an Eclipse project wizard allows easy creation of algorithms,
which is mainly used for teaching purposes.

\subsection{Lessons Learned}
\label{sec:lessons-learned}

In the following, we describe what users want from a layout library based on feature requests and questions from the \ac{elk} community.
We reflect on the maintenance and documentation practices of \ac{elk}, and present features that should be removed or reworked to foster \acposs{elk} future success.

\subsubsection{Listening to Users}

We interpret the success of \ac{elk} (and other layout libraries) to indicate that the idea of automatic layout is slowly but surely gaining ground in the practical usage of graphical languages.
However, to be truly successful, one still has to listen to the users.
Specifically, in our experience, users generally seem to be quite pleased with the layout results, at least as far as the aesthetic criteria are concerned that drawing algorithms are typically evaluated with, such as reducing edge crossings.
However, users may get frustrated if they appear to have \enquote{little control} over actual results.
When somebody's mental map of a graph is inflicted because drawings change drastically after minimal changes of the graph, or, even worse, without any changes of the graph, this also causes irritations.
For quite some time, we experimented with different approaches to give the users more control and stability.
For example, we provide annotations to the graph that specify the nodes that should be placed in the same layer \cite{PetzoldDSvH23}.
However, these did not really seem to catch on, probably for lack of awareness of these options. 

We, therefore, lately focused on a different approach to give modelers control, and to enhance layout stability as well, which is the \emph{model order} approach mentioned earlier.
Basically, we consider it---in most cases---a mistake to consider a graph as a \emph{set} of nodes and \emph{set} of edges, as done in most of the literature.
In our experience, it is much more practical to treat nodes and edges as ordered lists, with the order stemming \eg from a textual input file.
The model order should only be violated if the graph can clearly be improved by it.
Whenever there are equally good alternative drawings available, one should use model order as a \enquote{tie-breaker}, \ie, choose the drawing that is closer to the model order \cite{DomroesRvH23}.

For the same reasoning, while we originally used randomness rather liberally in our algorithms, in the hope to increase the chances of optimizing drawings according to some aesthetic criteria, we by now try to avoid randomness unless there are very strong reasons to include it.

\subsubsection{Maintaining and Documenting the Project}

Since its beginnings, \ac{elk} typically had (only) one to three active developers.
Furthermore, \ac{elk} is research driven, and whatever software engineering and maintenance it receives is mostly provided by motivated graduate students \enquote{on the side.}
Software engineering tasks such as adding documentation, fixing bugs,
testing, updating versions, interacting with users, and managing builds and releases cannot be the main focus of researchers, which mainly focus on new features and strategies for \ac{elk} for their dissertation.
At the same time, in computer science it is not common to fund software engineers that operate and maintain open-source libraries such as \ac{elk} by the research institution itself.
On the long run this is a severe impediment for research that builds on powerful, reliable open-source platforms.
The situation is different in other fields.
\Eg, in experimental physics it is understood that a research institution not only needs scientists that actually conduct and publish research, but that there are also highly skilled lab engineers required that tend to the infrastructure.
Given these circumstances, we originally would not have expected \acposs{elk} longevity and popularity.
\Eg, elkjs has received over 1200 \enquote{stars} on GitHub by now.

This maintenance effort also enables research.
Even though the first contributors to \ac{elk} have graduated long ago, \ac{elk}
still provides a stable research infrastructure for complex layout algorithms.
\Eg, the recently added model order strategies require a highly modular layout algorithm with exchangeable layout strategies for the Sugiyama algorithm and a flexible system for pre- and post-processing, which is provided by the phase and processor infrastructure of \ac{elk} presented in \autoref{fig:processor-slots}.

As part of the maintenance effort, early \ac{elk} developers lived and enforced good coding conventions consisting of regular code reviews and refactoring as part of the release cycle.
These made sure that the algorithm structure did not degrade over time.
Most of the releases in \autoref{fig:timeline} were accompanied by a refactoring process, especially for the initial release of \ac{klay} Layered, \ac{elk}, and elkjs.
\Eg, as part of the creation of \ac{elk} the graph format was customized for layout and anything regarding visualization was removed.
Moreover, the transpilation to elkjs required elimination of \enquote{hacks} such as Java reflection.

The move to the Eclipse Foundation was a time-consuming effort but finally helped to support the maintenance, documentation, and legal efforts.
As a result, \ac{elk} has a stable release and build management system, which does not depend on sometimes unreliable university server infrastructure.
Release reviews guarantee a certain code, documentation, and process quality, and the Eclipse Contributor Agreement prevents legal issues.

Additionally, documentation is partly enforced by convention and syntax.
The metadata language that describes algorithms configurations enforces documentation, which is hosted on the Eclipse project web-page together with additional documentation in form of guides, examples, and blog posts.
However, as in many academic projects, the end user documentation lacks behind.
Interactive support channels proved necessary.
As a result, \ac{elk} utilizes the online \ac{elk} editor elklive (see \autoref{sec:software}) and the \ac{elk} Gitter chat\footnote{\url{https://gitter.im/eclipse/elk}}, next to the GitHub issues to communicate with its user base.

Not only the end users but also developers need documentation.
For the layered algorithm, processors are carefully ordered in each slot based on dependencies between, which corresponds to the processor ordering in \autoref{fig:processor-all}.
Each processor documentation lists its designated slots, pre- and post-conditions, and in-slot dependencies to other processors in form of Javadoc.
However, the documentation of dependencies is not always done carefully since dependencies between processors and between different layout options get exponentially more complex with the number of features.
Moreover, new features might require constraints that were previously deemed irrelevant.
\Eg, the model order strategies might reorder nodes and edges and assume that other processors will not introduce additional nodes and will respect this ordering.
This was originally not considered since a graph was previously assumed to consist of \emph{unordered} sets of nodes and edges.

\subsubsection{Removing Outdated Features}

The metadata-compiler, which compiles the algorithm metadata into Java and generates algorithm and layout option documentation, on the one hand makes it easier for \ac{elk} developers to add new layout options.
On the other hand it requires an installed metadata-compiler in the development environment, which is, hence, limited to the Eclipse \ac{ide}.
In its current state, this limits collaboration in the open-source project.
We experience that highly motivated users want to collaborate but fail to install the necessary tooling.
At the same time, the metadata-compiler generates most of the algorithm documentation.
Reworking this would require an effort but would greatly help \ac{elk} gain additional contributors.

Debugging tools, which proved to be essential for algorithm development, only exist for the Eclipse \ac{ide}.
As part of future work the Eclipse-specific functions as well as part of the service architecture should move in a separate repository, leaving only the algorithms and their infrastructure in the \ac{elk} core project.

Refactoring should not only be used to improve code quality and to identify bottlenecks but also to remove features.
Some features, such as the diploma thesis of Fuhrmann (see \autoref{fig:timeline}) about compound graph layout, were too complex to be maintained or do not fit into the algorithm phase structure.
Such features may threaten the whole layout library since they increase the maintenance effort and facilitate the usage of \enquote{hacks} to build desired functionality.
Hence, regular refactoring needs to identify such features and deprecate them to decrease unnecessary maintenance efforts.

\section{Examples}
\label{sec:examples}

In the following, we showcase how \ac{elk} is used to layout SCCharts \cite{vonHanxledenDM+14}, a synchronous language that models control-flow, and Lingua Franca \cite{LohstrohMBL21}, a polyglot coordination language that models data-flow.
All the following models are styled with KLighD \cite{SchneiderSvH13}, since \ac{elk} does not support colors, fonts, and non-rectangular forms but only computes coordinates and sizes for rectangles and routes lines.

\begin{figure}
	\centering
	\includegraphics[scale=0.5]{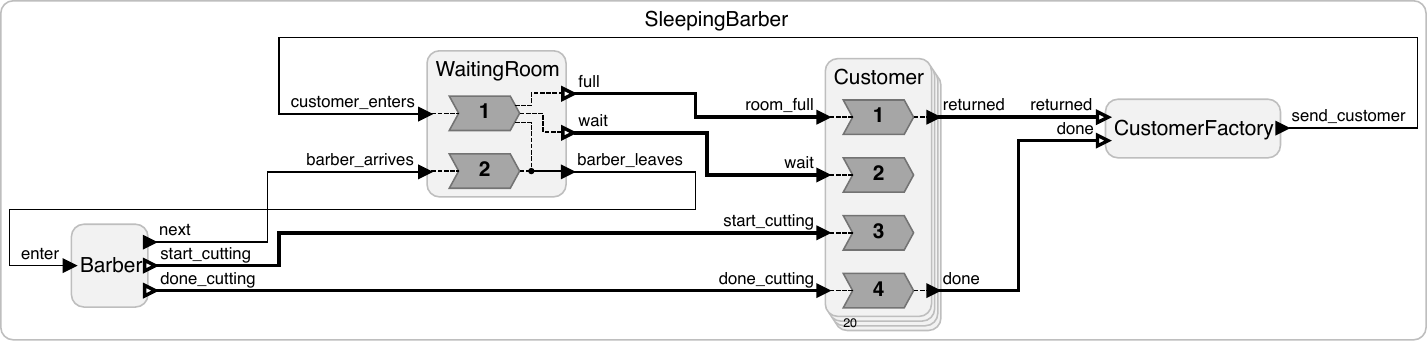}
	\caption{The sleeping barber Lingua Franca model taken from \url{https://github.com/lf-lang/examples-lingua-franca}.} 
	\label{fig:SleepingBarber}
\end{figure}

The Lingua Franca model in \autoref{fig:SleepingBarber} showcases hierarchical edges using \ac{elk} Layered.
The reactors \textsff{WaitingRoom} and \textsff{Customer} both have internal behavior, which is also layouted using \ac{elk} Layered.
Here, model order orders the internal behavior of \textsff{WaitingRoom} and \textsff{Customer} as their numbering suggests.
The inner graph, the ports, and the port labels determine the size of a compound node.
\ac{elk} Layered is here configured to place the port labels consistently above the corresponding port.
The placement of the vertical edge segments between \textsff{WaitingRoom} and \textsff{Customer} minimizes the used horizontal space.
The port-label spacing and port-port spacing makes sure that the labels does not overlap with the arrow shaped ports and that no ports overlap.

\begin{figure}
	\centering
	\includegraphics[scale=0.365]{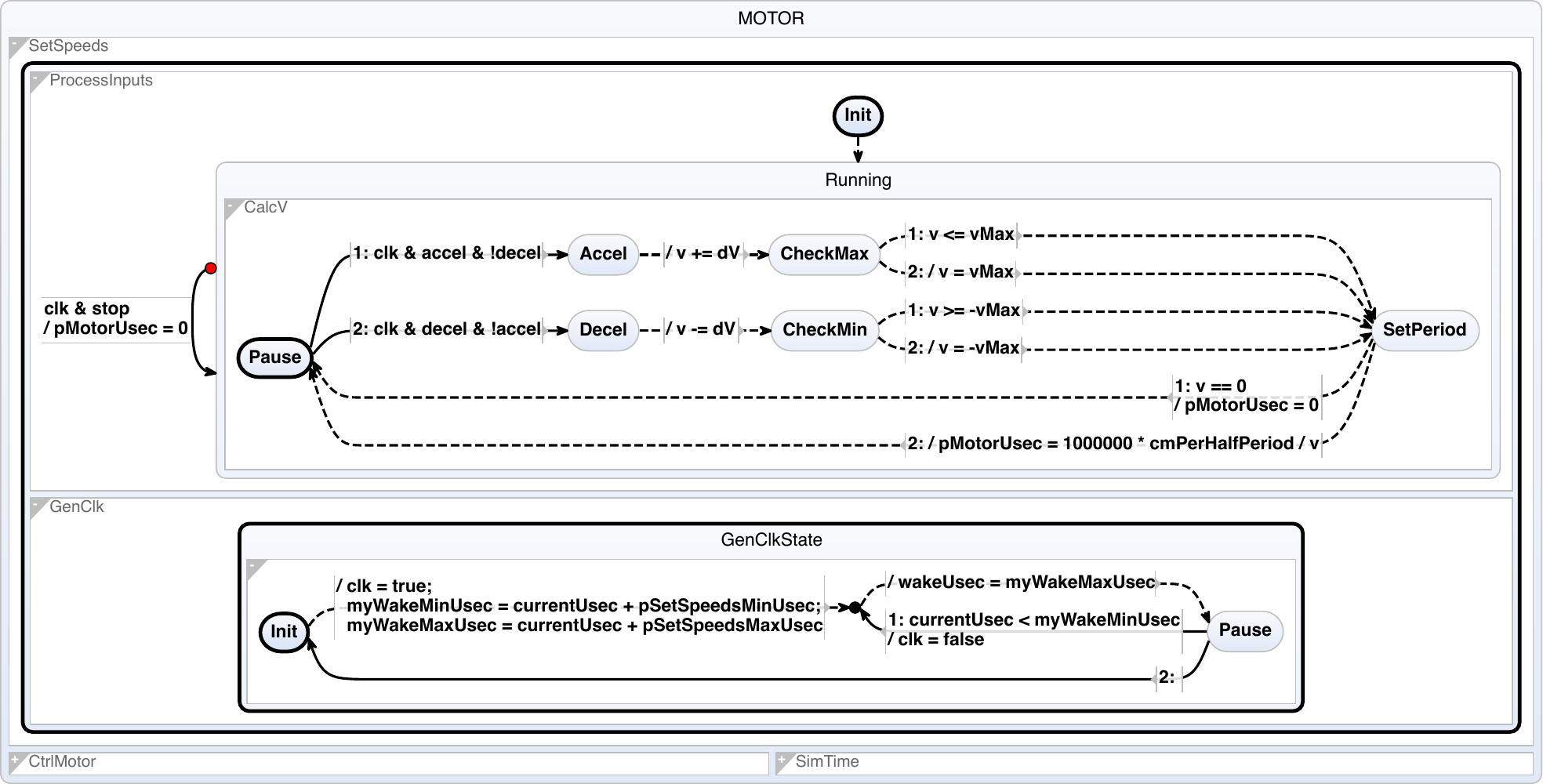}
	\caption{The motor SCChart.}
	\label{fig:motor}
\end{figure}

The SCChart in \autoref{fig:motor} showcases how different layout algorithms can be used in the same model.
Here, rectangle packings and layered layouts alternate.
Moreover, directions of layered layouts also alternate, creating drawings with a better aspect ratio.
The graph inside \textsff{ProcessInputs} is layouted top-to-bottom, and the graph inside \textsff{Running} and \textsff{GenClkState} is layouted left-to-right.
Model order is used as a tie-breaker, which makes sure that the outgoing and incoming edges of the initial \textsff{Pause} state are correctly ordered.
Model order also highlights the control-flow loops, since the textual state ordering inside an SCChart usually determines the choice for a backward edge.
Edge-label-management positions edge labels such that they have semantic line breaks.
Moreover, edge labels are positioned on an edge such that labels can be clearly matched with the transitions they refer to.
The \enquote{on-edge} positioning tends to reduce the drawing height.
Padding makes sure that the inner layered graph will not overlap with the triangle in the upper left corner or the label next to it.
In a corresponding \ac{ide}, the layout direction, label shortening strategies, and other layout options can be directly configured, \eg, to get a drawing to fit a certain aspect ratio or to create a desired layout.

The layout algorithm and all mentioned strategies are suitable in interactive scenarios and the layout time is for all shown models in millisecond range.

\section{Conclusion and Outlook}
\label{sec:conclusion}

\ac{elk} and the algorithms behind it are used as the basis of several academic and commercial projects, while only being developed by a handful of people over the course of nearly fifteen years.
Through its migration into the Eclipse Foundation and focusing not only on individual research but also on software engineering, it will hopefully live on for some time.

Graduate students not only spent a considerable amount of time engineering \ac{elk} but also implemented good coding and documentation practices to maintain a high code quality.
Time spent interacting with the community through multiple interactive collaboration channels allows identifying highly requested features and developing practically relevant algorithms.
These new use cases spawn new research topics, which again attracts future developers for \ac{elk} in form of graduate students.
However, \ac{elk} could greatly benefit from a non-scientific software engineer.

Future work on \ac{elk} will include refactoring as well as general maintenance issues to make \ac{elk} more robust for future development.
\ac{elk} will add support for top-down layout as an alternative to the presented bottom-up layout strategy for compound graphs, which can be used to add map-navigation-like tooling for complex graphs.
Moreover, \ac{elk} Layered will get support for selective in-layer edges, which are usually forbidden in layered layouts.
We also plan to add more model order configurations as well as model order support using different model order groups used to discern the ordering of different kind of graph elements.

\section{Acknowledgments}

We thank all the developers that put their heart and soul into \ac{elk}, which are partly mentioned in \autoref{fig:timeline}.
We are indebted to the numerous users of ELK providing valuable feedback.
In particular, Edward Lee and his group have been enthusiastic and supportive for many years now, which helped to considerably broaden the scope and utility of \ac{elk}.

\nocite{SpoenemannFvH09}
\nocite{SpoenemannFvH+09}
\nocite{KlauskeSS+12}
\nocite{GutwengervHM+14}
\nocite{SchulzeSvH14}
\nocite{SpoenemannDvH14b}
\nocite{SpoenemannSRvH14}
\nocite{RueeggSCvH15}
\nocite{RueeggLPK+16}
\nocite{JabrayilovMMR+16}
\nocite{RueeggSGvH16b}
\nocite{RueeggESvH16}
\nocite{SchulzeLvH16}
\nocite{Schulze16}
\nocite{RueeggvH18a}
\nocite{SchulzeHvH18a}
\nocite{SchulzeWvH18a}
\nocite{DomroesLvHJ21}
\nocite{DomroesvH22}
\nocite{DomroesLvHJ23}
\nocite{DomroesRvH23}
\nocite{PetzoldDSvH23}

\nocite{Spoenemann15}
\nocite{Rueegg18}
\nocite{Schulze19}
\bibliographystyle{splncs04}
\bibliography{../bib/cau-rt,../bib/pub-rts,../bib/rts-arbeiten,gd23.bib}
\end{document}